\documentclass[final,5p,times,twocolumn,authoryear]{elsarticle}

\usepackage[authoryear]{natbib}
\usepackage{graphicx}
\usepackage{subcaption}
\usepackage{amssymb,amsmath}
\usepackage{pdflscape}
\usepackage{adjustbox}
\usepackage{hyperref}
\usepackage{booktabs}
\usepackage{lipsum}
\usepackage{epsfig}
\usepackage{txfonts}
\usepackage{xcolor}
\usepackage{tabularx}
\usepackage{footnote}
\usepackage{soul}
\usepackage{multirow}
\usepackage{multicol}
\usepackage{mathpazo}
\usepackage{lineno,hyperref}
\usepackage{textcomp}
\usepackage{url}
\usepackage[utf8]{inputenc}
\usepackage{booktabs}

\usepackage{amsthm}

\journal{High Energy Astrophysics}

\begin{document}

\begin{frontmatter}



\title{Spectral State Switching in Mrk 421: Results from the  AstroSat LAXPC/SXT Observations}


%

\author{Sikandar Akbar\corref{cor1}\fnref{label1}}
\ead{darprince46@gmail.com}
\affiliation[label1]{Department of Physics, University of Kashmir, Srinagar 190006, India}
\cortext[cor1]{Corresponding author: Sikandar Akbar}

\author{Zahir Shah\corref{cor1}\fnref{label2}}
\ead{shahzahir4@gmail.com}
\affiliation[label2]{Department of Physics, Central University of Kashmir, Ganderbal 191201, India.}

\cortext[cor1]{Corresponding author: Zahir Shah}

\author{Ranjeev Misra\fnref{label3}}
\affiliation[label3]{Inter-University Centre for Astronomy and Astrophysics,  Post Bag  4, Ganeshkhind, Pune-411007, India}

\author{Naseer Iqbal\fnref{label1}}


\begin{abstract}
We carried a detailed time and flux resolved  X-ray spectral analysis of the high-synchrotron-peaked blazar Mrk\,421 using simultaneous LAXPC20 and SXT observations. The 100\,s binned LAXPC20 light curve obtained during 3--8 January 2017 reveals pronounced flux variability. The source exhibits a fractional variability amplitude of $F_{\mathrm{rms}} = 0.210 \pm 0.005$ in the SXT band and $F_{\mathrm{rms}} = 0.316 \pm 0.006$ in the LAXPC20 band. During this interval, the source reached a peak LAXPC20 count rate of
122.94\,counts\,s$^{-1}$, while the peak count rate in the SXT light curve is  26.79\,counts\,s$^{-1}$.  This enabled us to carry out flux-resolved spectroscopy by dividing the  100\,s binned LAXPC20 light curve into ten flux states (S1--S10), each spanning a width of 8\,counts\,s$^{-1}$. For each flux state, simultaneous SXT and LAXPC20 spectra were extracted and fitted jointly. We find that the spectra in these states are well described by a synchrotron-convolved broken power-law, which provides a better fit than a log-parabola model. The low-energy particle index (index before the break) is found to cluster around two discrete values across flux states indicating two spectra states in the source. The break energy consistently moves to high energy with increase in flux level in these states.  Time-resolved spectroscopy (10-ks segments) confirms that the flux histogram is best modelled as a double lognormal distribution and the index histogram is double normal. Inclusion of two additional long observations spanning 2017-2019 shows the same double-state behaviour on longer timescales. Together, the results indicate that Mrk\,421 routinely occupies two dominant spectral; in a leptonic synchrotron framework this can be explained by Gaussian-like fluctuations in acceleration conditions producing lognormal flux states.
\end{abstract}


\begin{keyword}
galaxies: active – BL Lacertae objects: general – BL Lacertae objects: individual: Mrk\,421 – galaxies: jets –
X-rays: galaxies.

\end{keyword}

\end{frontmatter}




\label{introduction}

Blazars exhibit strong and stochastic flux variability across the entire electromagnetic spectrum, spanning time-scales from minutes to years \citep{1997ARA&A..35..445U, bhatta2020nature, bhatta2016multifrequency}. 
Investigating the statistical properties of this variability provides key insights into the underlying emission processes. 
The form of the long-term flux distribution is particularly informative: a normal  distribution generally indicates additive and linear stochastic variability, whereas a log-normal distribution reflects non-linear, multiplicative processes. 
Such behaviour has been identified in several astrophysical systems, including X-ray binaries, gamma-ray bursts, and active galactic nuclei (AGNs) \citep{lognorm_xrb, 2002PASJ...54L..69N, 1997MNRAS.292..679L, 2002A&A...385..377Q, 2009A&A...503..797G, 2018RAA....18..141S, 2020MNRAS.496.3348S}.

The variability observed in AGNs occurs on time scales from minutes to days \citep{2004ApJ...612L..21G} and in  X-ray binaries \citep{2005MNRAS.359..345U}, it occurs on  sub-second time scales. The emission in these sources  occurs from the accretion discs and thus observation of log-normal distribution of flux variations in these sources implies that underlying physical process are multiplicative in nature.
The first detection of a log-normal  flux distribution in a blazar was reported for BL\,Lacertae \citep{2009A&A...503..797G}, hinting at a possible link between accretion-driven fluctuations and jet emission. 
Subsequent observations of blazars such as PKS~2155$-$304 and Mrk~501 have confirmed similar patterns in both X-ray and very-high-energy (VHE; $>$100\,GeV) $\gamma$-ray bands \citep{2009A&A...502..749A, 2010A&A...520A..83H, 2010A&A...524A..48T, 2018Galax...6..135R, 2019MNRAS.484..749C}. 
Long-term Fermi-LAT observations also reveal log-normal $\gamma$-ray flux distributions in several bright blazars, including the VHE source 1ES~1011+496 \citep{2018RAA....18..141S, my1011}. 
In some sources, double log-normal or composite distributions provide better fits, indicating multiple emission states or regions within the jet \citep{pankaj_ln, 2018RAA....18..141S, 2016ApJ...822L..13K, 2019MNRAS.484.3168S, 2020MNRAS.491.1934K}.

The origin of log-normal variability is often attributed to multiplicative processes within the accretion flow \citep{2005MNRAS.359..345U, 2010LNP...794..203M}. 
However, disk fluctuations alone cannot explain the minute-scale variations observed in many blazars \citep{1996Natur.383..319G, vhe501, vaidehi421}. 
Alternative explanations invoke jet-based mechanisms such as emission from numerous randomly oriented mini-jets within the main relativistic jet \citep{minijet}, or stochastic perturbations in the particle acceleration timescale that produce Gaussian variations in the particle index and consequently a log-normal flux distribution \citep{2018MNRAS.480L.116S}. 
These processes highlight the importance of turbulence, shocks, and particle cascades within relativistic jets \citep{2010MNRAS.402..497G}, and the observed correlation between  index and flux further supports a direct coupling between particle acceleration and the evolving energy distribution in the jet \citep{2010ApJ...716...30A, 2021MNRAS.508.5921H}.

Mrk\,421, located at a redshift $z \approx 0.031$, is one of the nearest and brightest VHE blazars in the extragalactic sky. 
It was first detected in $\gamma$-rays by the Energetic Gamma Ray Experiment Telescope \citep[EGRET;][]{lin1992detection} and was also the first confirmed TeV blazar observed by the Whipple telescope \citep{punch1992detection, petry1996detection}. 
It has since been extensively studied across the electromagnetic spectrum \citep{shukla2012multiwavelength, sinha2016long, krawczynski2001simultaneous}, showing pronounced variability in X-ray and $\gamma$-ray bands on time-scales from minutes to days \citep{doi:10.1146/annurev.aa.33.090195.001115, falomo2014optical, goyal2020blazar}. 
The rapid X-ray variability indicates that the radiating electrons are highly energetic and have short cooling times \citep{2007ApJ...664L..71A, ackermann2016minute}.

Building on these findings, \citet{2021MNRAS.508.5921H, AKBAR2025438} carried out a detailed time-resolved X-ray spectral analysis of Mrk\,421 using multiple AstroSat observations between 2016 and 2019. \citet{2021MNRAS.508.5921H} showed that the spectrally degenerate models, can be distinguished based on spectral parameter correlations.
 \citet{AKBAR2025438} found a constant pivot energy across all epochs, a result that remained model-independent for the broken power-law (BPL), log-parabola (LP), and power-law with maximum electron energy ($\xi_{\text{max}}$) representations of the particle energy distribution. 
The invariance of pivot energy suggests that the X-ray variability in Mrk\,421 is primarily driven by changes in the spectral index rather than normalization, implying that parameters such as magnetic field strength and Doppler factor remain relatively stable. 
Therefore, the dominant source of variability is most likely related to fluctuations in the particle acceleration or escape timescales within the jet emission region.

Previous studies have reported two-state spectral behaviour in Mrk 421. However, it has remained unclear whether such two-state behaviour reflects an intrinsic physical property of the source or depends on the specific analysis methodology. In this work, we demonstrate that the two-state behaviour emerges independently from flux-resolved spectroscopy and from time-resolved statistical distributions, indicating that it is not driven by methodological choices alone. We investigate the flux and spectral index distribution properties of Mrk 421 using flux- and time-resolved X-ray spectral analysis of AstroSat observations, and discuss the implications for particle acceleration and spectral variability in jet-dominated blazars.

 The paper is organized as follows: Section \ref{reduction} covers data reduction, Section \ref{analysis} presents the results of flux and time-resolved spectral analysis, and a detailed correlation study. Section \ref{summary} offers a summary of our work, followed by an in-depth discussion.

\section{DATA  REDUCTION }\label{reduction}

AstroSat, India’s first dedicated multiwavelength space observatory, offers the distinctive capability of \emph{simultaneous observations spanning optical/UV to soft and hard X-ray energies} \citep{2017JApA...38...27A}. The mission carries five scientific payloads, including the Ultra-Violet Imaging Telescope (UVIT), which operates in the 130--300\,nm band \citep{2017JApA...38...28T,2017AJ....154..128T}; the Soft X-ray Telescope (SXT), sensitive to the 0.3--8.0\,keV energy range \citep{2016SPIE.9905E..1ES,2017JApA...38...29S}; the Large Area X-ray Proportional Counter (LAXPC), covering energies from 3 to 80\,keV \citep{2016SPIE.9905E..1DY}; and the Cadmium Zinc Telluride Imager (CZTI), operating in the 10--100\,keV band \citep{2017CSci..113..595R}.

AstroSat observed Mrk\,421 during During the period 2017-2019, employing both the SXT and LAXPC20 instruments. All data were retrieved from the AstroSat archive through the ASTROBROWSE interface\footnote{\url{https://astrobrowse.issdc.gov.in/astro_archive/archive/Home.jsp}}. Details of the data reduction and analysis procedures for the SXT and LAXPC observations are described in the following sections.

\subsection{Soft X-ray Telescope (SXT)}
The Soft X-ray Telescope (SXT) onboard AstroSat is a grazing-incidence X-ray imaging instrument with a focal length of 2\,m, operating over the 0.3--8.0\,keV energy band \citep{2017JApA...38...29S}. It employs an e2V CCD-22 detector positioned at the common focal plane of the mirror assembly. The telescope provides an angular resolution of approximately 2\,arcmin and a circular field of view with a diameter of $\sim$40\,arcmin.

SXT observed Mrk\,421 in photon counting (PC) mode during multiple epochs between 2017-2019. The Level-1 SXT data were processed using the standard SXT data reduction pipeline version~1.4b (AS1SXTLevel2-1.4b; released on 2019 January 03). Event files from individual orbits were subsequently merged using the \texttt{SXTEVTMERGER} tool. Standard science products, including light curves, images, and spectra, were extracted using the \texttt{XSELECT} tool (version~2.4m) distributed with \texttt{HEASOFT}~v6.29.
For source extraction, a circular region of radius 16\,arcmin centered on the source position was used, enclosing more than 95\% of the detected source photons. An off-axis auxiliary response file (ARF), appropriate for the chosen extraction region, was generated using the \texttt{sxtARFModule}. The background spectrum \texttt{SkyBkg\_comb\_EL3p5\_Cl\_Rd16p0\_v01.pha}, provided by the SXT Payload Operations Centre (POC), along with the response matrix file \texttt{sxt\_pc\_mat\_g0to12.rmf}, were employed for spectral analysis. To ensure adequate photon statistics in each spectral bin, the spectra were grouped using the \texttt{ftgrouppha} command.

 \subsection{Large Area Proportional Counters
(LAXPC)}

The Large Area X-ray Proportional Counter (LAXPC) is one of the primary payloads onboard AstroSat. It is a non-focusing X-ray instrument with a large combined effective area of $\sim$6000\,cm$^{2}$. The instrument consists of three identical proportional counter units, namely LAXPC10, LAXPC20, and LAXPC30. These detectors operate over the 3.0--80.0\,keV energy range and provide a high temporal resolution of 10\,$\mu$s \citep{2016SPIE.9905E..1DY,2017ApJS..231...10A,2017JApA...38...30A,2017ApJ...835..195M}.
In this work, we exclusively used data from LAXPC20. The LAXPC30 unit was switched off in March 2018 due to abnormal gain variations \citep{2017ApJS..231...10A}, while the application of standard response files to LAXPC10 was not feasible because of changes in the high-voltage settings during the spring of 2018. Consequently, LAXPC20 provided the most reliable and stable dataset for the present analysis.

The data reduction and analysis were carried out using the \texttt{LAXPCSOFT} software package (version \texttt{LAXPCsoftware22Aug15}). A Level-2 event file was first generated using the task \texttt{laxpc\_make\_event}. Subsequently, a good time interval (GTI) file was created with \texttt{laxpc\_make\_stdgti} to exclude periods affected by South Atlantic Anomaly (SAA) passages and Earth occultation. The resulting GTI file was then applied to extract the source light curves and spectra.
Although Mrk\,421 is bright in the 3--30\,keV band, it becomes
background dominated in the 50--80\,keV band. This allows the
use of the faint-source background estimation scheme implemented
in \texttt{LAXPCSOFT} to generate background light curves and
spectra for LAXPC20 \citep{misra2021alternative}. Background subtraction was performed using the standard \texttt{FTools} task \texttt{lcmath}. Owing to the increasing dominance of background at higher energies, we restricted the spectral analysis to the 3--20\,keV energy range.

\section{ANALYSIS}\label{analysis}
\subsection{Time variability}
We generated the X-ray light curves of Mrk\,421 during 3--8 January 2017 
(Observation ID: A02\_005T01\_9000000948). 
The LAXPC20 light curve was extracted in the 3.0--30.0\,keV energy range, 
while the SXT light curve covers the 0.8--7.0\,keV band. 
\textbf{The 100\,s binned background-subtracted LAXPC20 and SXT light curves are shown in Figure~\ref{fig:lightcurve}. }
We evaluated the variability characteristics of the X-ray light curves using the 
FTOOL \texttt{LCSTATS}. This task provides the fractional root-mean-square 
variability ($F_{\mathrm{rms}}$), which we determined from the 100\,s binned 
SXT and LAXPC20 light curves. For this observation, the fractional variability 
amplitude is $F_{\mathrm{rms}} = 0.2103 \pm 0.0051$ and 
$0.3159 \pm 0.0056$ for SXT and LAXPC20, respectively. 
These values indicate that the source exhibited significant variability during 
this period. 
The peak count rate reached $122.94$\,counts\,s$^{-1}$ in the LAXPC20 light curve, 
while the  peak count rate in the SXT light curve is 
 26.79\,counts\,s$^{-1}$.

\subsection{Flux resolved spectroscopy}

\begin{figure}
\centering

\begin{subfigure}{\linewidth}
\includegraphics[width=\linewidth]{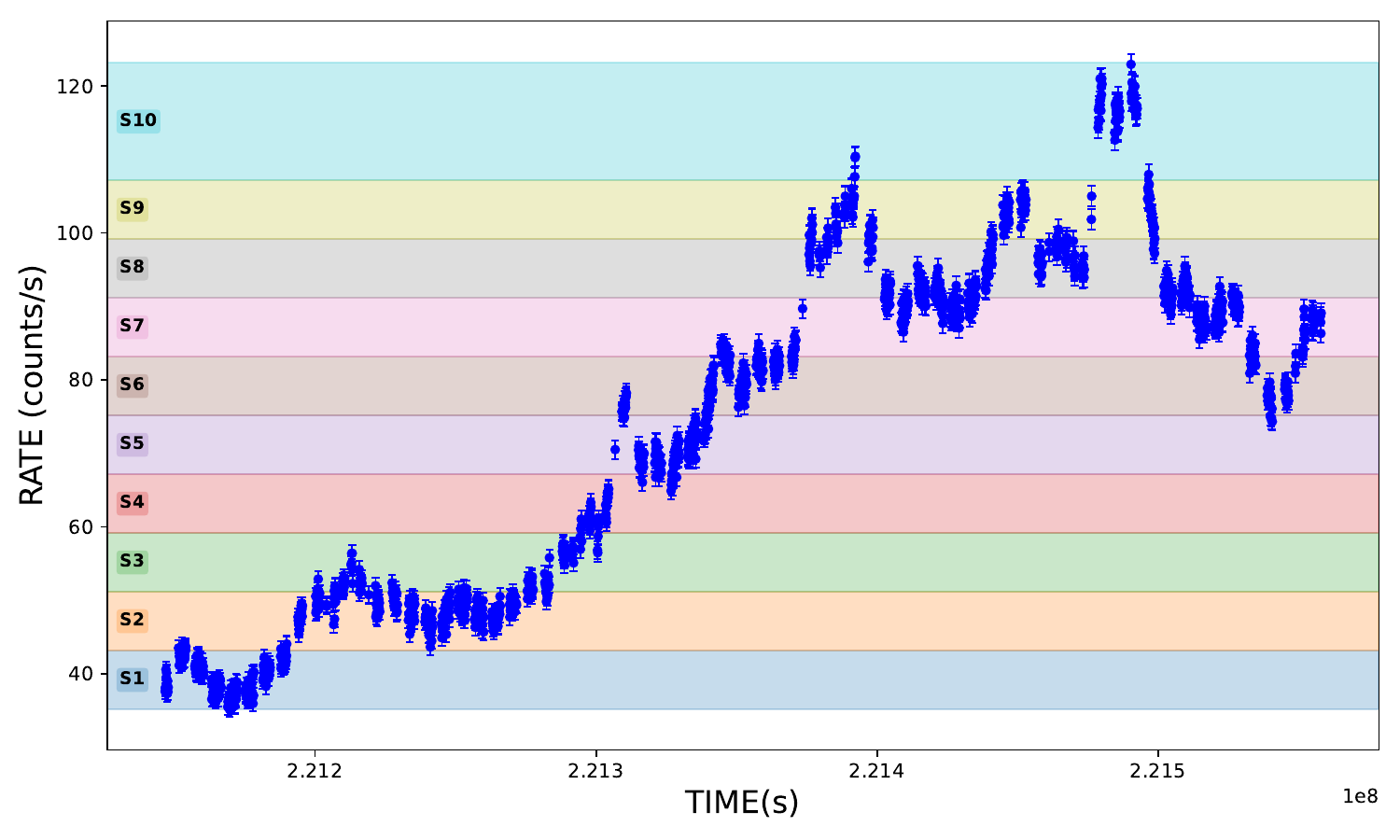}
\caption{LAXPC20 light curve (3.0--30.0\,keV)}
\end{subfigure}

\vspace{0.2cm}

\begin{subfigure}{\linewidth}
\includegraphics[width=\linewidth]{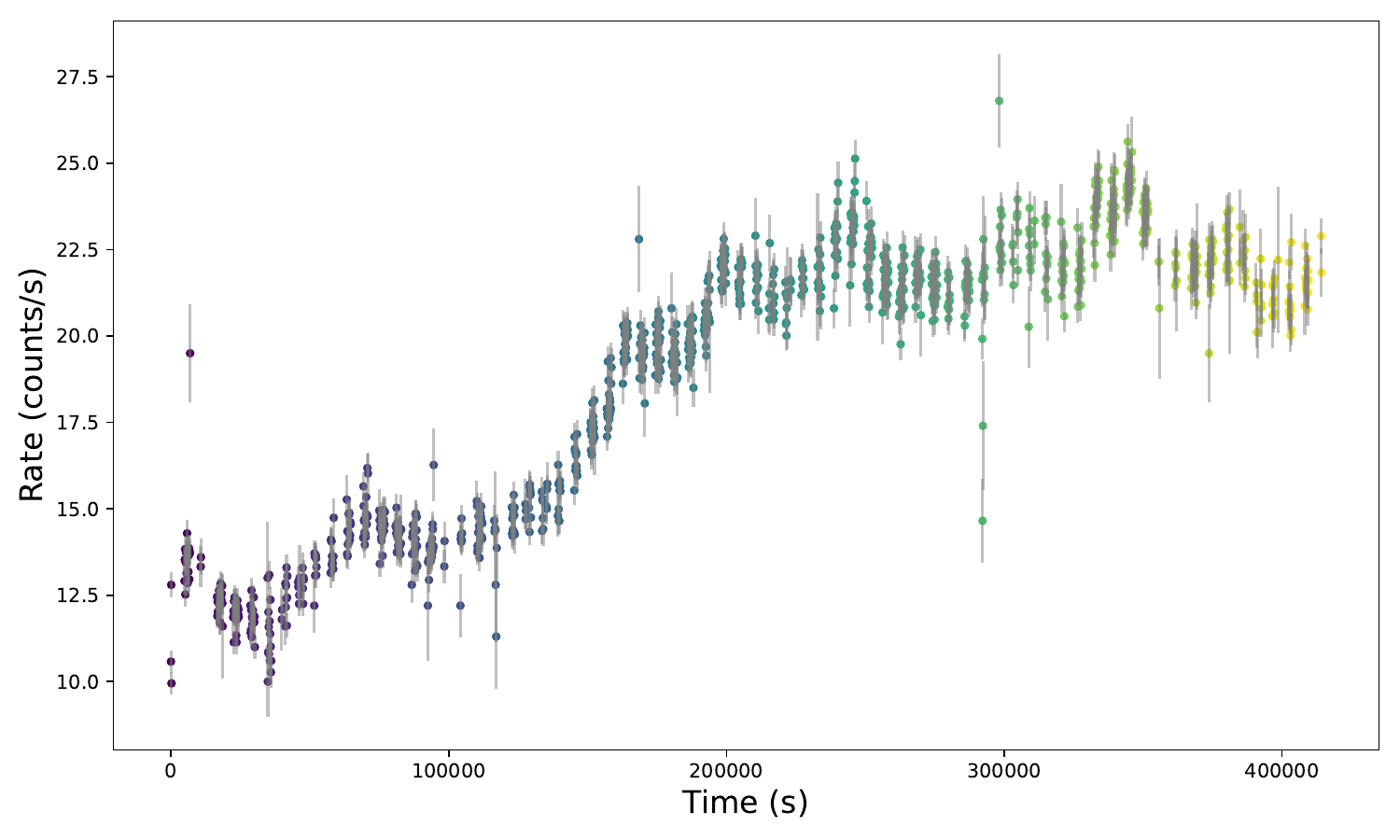}
\caption{SXT light curve (0.8--7.0\,keV)}
\end{subfigure}

\caption{
100\,s binned X-ray light curves of Mrk\,421 obtained during the
AstroSat observation (ID: A02\_005T01\_9000000948).
The top panel shows the LAXPC20 light curve in the 3.0--30.0\,keV band,
while the bottom panel shows the SXT light curve in the 0.8--7.0\,keV band.
The peak count rate of 122.94 counts\,s$^{-1}$ corresponds to the
LAXPC20 light curve. \textbf{Both the LAXPC20 and SXT light curves are background subtracted.}
}
\label{fig:lightcurve}
\end{figure}

To investigate the spectral behaviour of Mrk\,421 as a function of flux,
we performed flux-resolved spectroscopy using the background-subtracted
LAXPC20 light curve with 100\,s binning. These  light-curve data were then divided into flux
intervals of width 8\,counts\,s$^{-1}$. Each interval was assigned a
flux state (S1, S2, $\dots$), ordered from the lowest to the highest
flux. Since the number of data points at the highest flux levels was
comparatively smaller, the two uppermost flux intervals were merged to
improve the statistical quality of the spectra, resulting in a total of
ten flux states (S1--S10; see Figure~\ref{fig:lightcurve}).

For each flux state, a Good Time Interval (GTI) file was generated using
the 100\,s binned LAXPC20 light curve with the \texttt{laxpc\_fluxresl\_gti}
command, which identifies time intervals where the count rate lies
between specified minimum (\texttt{-l}) and maximum (\texttt{-h}) values:
\begin{verbatim}
laxpc_fluxresl_gti -l ratemin -h ratemax \
-o output_gti_filename filename.lc
\end{verbatim}

The resulting GTI files were then used to extract simultaneous SXT and
LAXPC20 spectra corresponding to each flux state, employing the same
response and background files as used in the time-averaged analysis.
These flux-resolved spectra were subsequently modelled to investigate
the evolution of the spectral parameters with source flux.

\subsection{X-ray spectral analysis}\label{xray_spectral}

We analyzed the X-ray spectra of Mrk\,421 using \textsc{xspec} version~12.12.0 in the 0.8--20\,keV energy range. 
The SXT data were used in the 0.8--7.0\,keV band, while the LAXPC20 spectra covered 3.0--20.0\,keV. 
For the SXT, the background was estimated using the file \texttt{SkyBkgcomb\_EL3p5\_Cl\_Rd16p0\_v01.pha}, 
whereas the LAXPC background spectra were generated using the faint source background model provided by the LAXPC team. 
A systematic error of 3\% was applied to all fits. Spectral grouping was performed using the ``\texttt{ftgrouppha}''
tool for the SXT spectra, adopting the optimal binning scheme
of \citet{refId0} through the ``\texttt{opt}'' option. This method determines the bin size using the instrumental response
to avoid oversampling of the detector resolution while retaining adequate statistical quality in each spectral bin. For the LAXPC
spectra, channels were grouped at the $\sim5\%$ level to obtain approximately three energy bins per detector resolution element,
consistent with the moderate energy resolution of LAXPC ($\sim10$--$14\%$; \citealt{2017ApJS..231...10A, 2017CSci..113..591Y}).

The X-ray emission from Mrk\,421 is attributed to synchrotron radiation produced by non-thermal relativistic electrons. 
We modeled the spectrum assuming that the emission arises from a homogeneous spherical region of radius~$R$, 
filled with a magnetic field~$B$ and an isotropic distribution of relativistic electrons $n(\gamma)$ undergoing synchrotron losses. 
The pitch-angle-averaged synchrotron power per unit frequency emitted by a single electron is given by \citep{1986rpa..book.....R}

\begin{equation}
    P(\gamma,\nu) = \frac{\sqrt{3}\pi e^{3} B}{4 m c^{2}} F\left(\frac{\nu}{\nu_{c}}\right),
\end{equation}

where $\nu_{c} = \frac{3\gamma^{2}eB}{16mc}$, and $F\left(\nu/\nu_{c}\right)$ is the synchrotron function defined as
\begin{equation}
    F(x) = x \int_{x}^{\infty} K_{5/3}(\psi)\, d\psi,
\end{equation}
with $K_{5/3}(\psi)$ representing the modified Bessel function of order~$5/3$.

The synchrotron emissivity from a distribution $n(\gamma)$ is then expressed as
\begin{equation}
    J_{\rm syn}(\omega,\alpha) = \frac{1}{4\pi} \int_{\gamma_{\rm min}}^{\gamma_{\rm max}} P(\gamma,\omega,\alpha) n(\gamma)\, d\gamma.
\end{equation}
By substituting $\xi = \gamma \sqrt{C}$, with $C = 1.36 \times 10^{-11} \frac{\delta B}{1+z}$ (where $z$ is the redshift and $\delta$ the Doppler factor), 
the synchrotron flux observed at energy $\epsilon$ can be written as \citep{begelman1984theory}

\begin{equation}\label{eq:syn_obs}
    F_{\rm syn}(\epsilon) = \frac{\delta^{3}(1+z)}{d_{L}^{2}}\, V\, \mathbb{A} 
    \int_{\xi_{\rm min}}^{\xi_{\rm max}} F\!\left(\frac{\epsilon}{\xi^{2}}\right) n(\xi)\, d\xi,
\end{equation}
where $\mathbb{A} = \frac{\sqrt{3}\pi e^{3} B}{16 m_{e} c^{2} \sqrt{C}}$, 
$V$ is the emitting volume, and $d_{L}$ denotes the luminosity distance.

\begin{figure}
    \centering
    \includegraphics[width=1.1\linewidth]{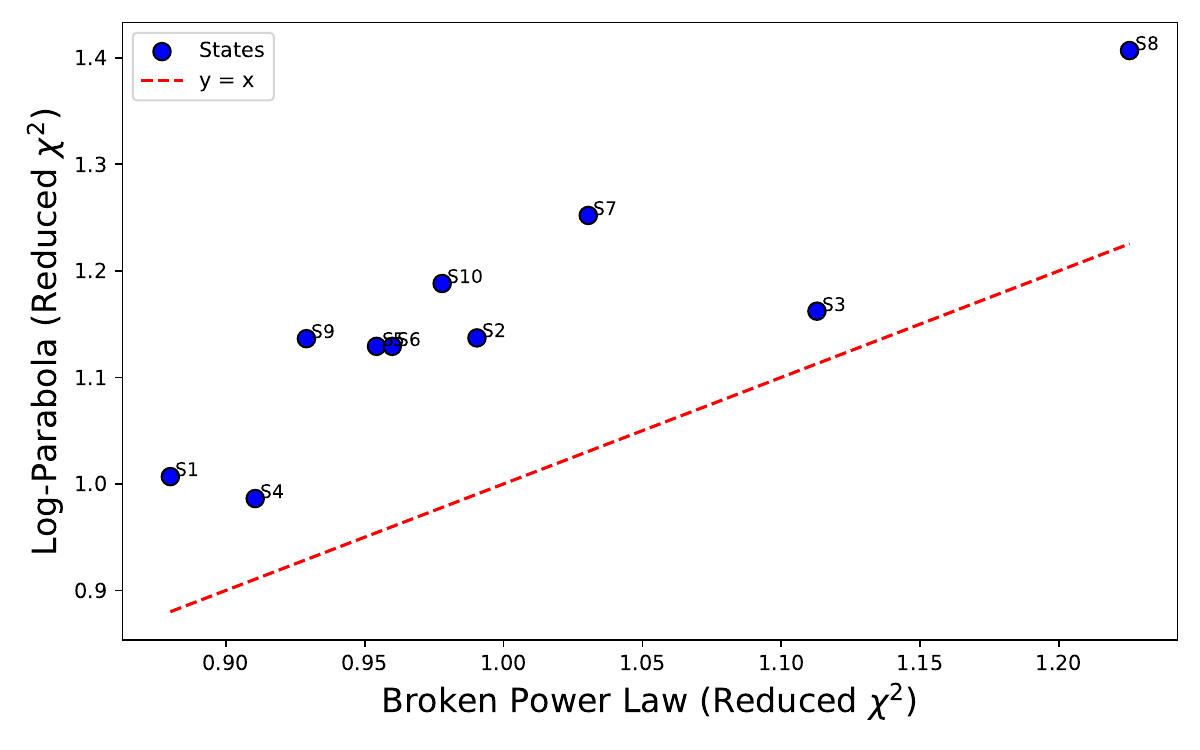}
    \caption{Comparison of the reduced chi-square ($\chi^2_\nu$) values obtained from the broken power-law and log-parabola fits for different spectral states of Mrk\,421. 
}
    \label{fig:chi}
\end{figure}

Equation~\ref{eq:syn_obs} was numerically solved and implemented in \textsc{xspec} as a local convolution model, 
\texttt{synconv~$\otimes$~n($\xi$)}, allowing the photon spectrum to be modeled for any assumed particle energy distribution~$n(\xi)$. 
In this model, the XSPEC variable \texttt{Energy} is defined as $\xi = \sqrt{C}\gamma$. 
We considered two forms for $n(\xi)$, a broken power law (BPL) and a log-parabola (LP) model, to fit the observed spectra. 
Joint fitting of the SXT and LAXPC20 spectra was performed using the model:
\[
\textit{constant} \times \textit{TBabs} \times (\textit{Synconv} \otimes n(\xi)).
\]
The hydrogen column density was fixed at $N_{\rm H} = 1.33 \times 10^{20}\, \text{cm}^{-2}$, 
as obtained from the LAB survey \citep{2005A&A...440..775K}. 
Galactic absorption was modeled using the \textsc{TBabs} routine. 
A relative normalization constant was applied to the SXT spectrum to account for inter-instrument calibration differences. 
During fitting, the \texttt{gain} command was employed with the slope fixed at 1, allowing the offset to vary. The broken power-law model provides a statistically better fit to the X-ray spectra compared to the log-parabola model (see Figure~\ref{fig:chi}). The resulting flux resolved spectral parameters are presented in Table \ref{tab:bkn_table}. 

\begin{table*}
\centering
\caption{Broken power-law spectral parameters for different flux states of Mrk\,421. 
Columns list the low- and high-energy  indices ($\Gamma_1$ and $\Gamma_2$), normalization (n, scaled by a factor of 100), break energy ($\xi_{\rm brk}$, in keV), and reduced chi-square ($\chi^2_\nu$). 
}
\label{tab:bkn_table}
\begin{tabular}{lccccc}
\hline
State & $\Gamma_1$ & $\Gamma_2$ & $n \times 100$ & $\xi_{\rm brk}$ & $\chi^2_\nu$ \\
\hline
S1  & $2.49^{+0.11}_{-0.11}$ & $4.29^{+0.08}_{-0.08}$ & $22.94^{+0.00}_{-0.00}$ & $1.53^{+0.14}_{-0.09}$ & 0.88 \\
S2  & $2.56^{+0.10}_{-0.08}$ & $4.13^{+0.09}_{-0.08}$ & $26.24^{+0.00}_{-0.00}$ & $1.75^{+0.32}_{-0.18}$ & 0.99 \\
S3  & $2.50^{+0.13}_{-0.14}$ & $4.19^{+0.09}_{-0.09}$ & $28.08^{+0.01}_{-0.01}$ & $1.58^{+0.28}_{-0.21}$ & 1.11 \\
S4  & $2.56^{+0.11}_{-0.11}$ & $4.56^{+0.13}_{-0.12}$ & $33.05^{+0.01}_{-0.01}$ & $1.85^{+0.37}_{-0.22}$ & 0.91 \\
S5  & $2.51^{+0.07}_{-0.07}$ & $4.43^{+0.10}_{-0.10}$ & $35.90^{+0.01}_{-0.01}$ & $1.81^{+0.22}_{-0.15}$ & 0.95 \\
S6  & $2.49^{+0.08}_{-0.08}$ & $4.42^{+0.10}_{-0.10}$ & $37.86^{+0.01}_{-0.01}$ & $1.86^{+0.28}_{-0.18}$ & 0.96 \\
S7  & $2.14^{+0.08}_{-0.08}$ & $4.40^{+0.09}_{-0.09}$ & $37.99^{+0.01}_{-0.01}$ & $1.78^{+0.09}_{-0.08}$ & 1.03 \\
S8  & $2.13^{+0.08}_{-0.09}$ & $4.11^{+0.08}_{-0.08}$ & $38.85^{+0.01}_{-0.01}$ & $1.88^{+0.30}_{-0.20}$ & 1.22 \\
S9  & $2.15^{+0.09}_{-0.09}$ & $4.09^{+0.09}_{-0.09}$ & $38.73^{+0.01}_{-0.01}$ & $1.85^{+0.26}_{-0.20}$ & 0.93 \\
S10 & $2.11^{+0.08}_{-0.08}$ & $4.07^{+0.10}_{-0.09}$ & $38.57^{+0.01}_{-0.01}$ & $2.00^{+0.38}_{-0.26}$ & 0.98 \\
\hline
\end{tabular}
\end{table*}

\begin{figure*}
    \centering
    \includegraphics[width=\linewidth]{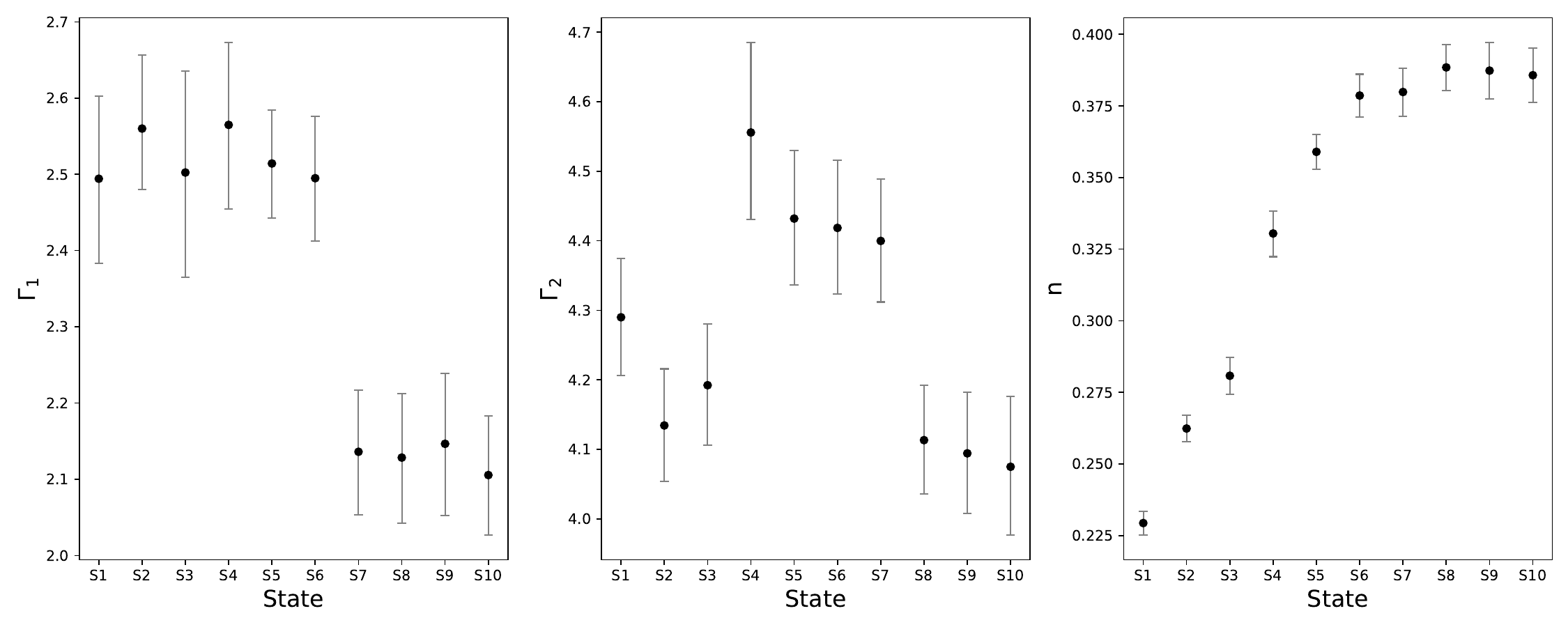}
    \caption{Variation of the broken power-law spectral parameters of Mrk\,421 with flux state (S1--S10).}
    \label{fig:bknpow_state_trends}
\end{figure*}

\begin{figure*}
    \centering
    \includegraphics[width=\linewidth]{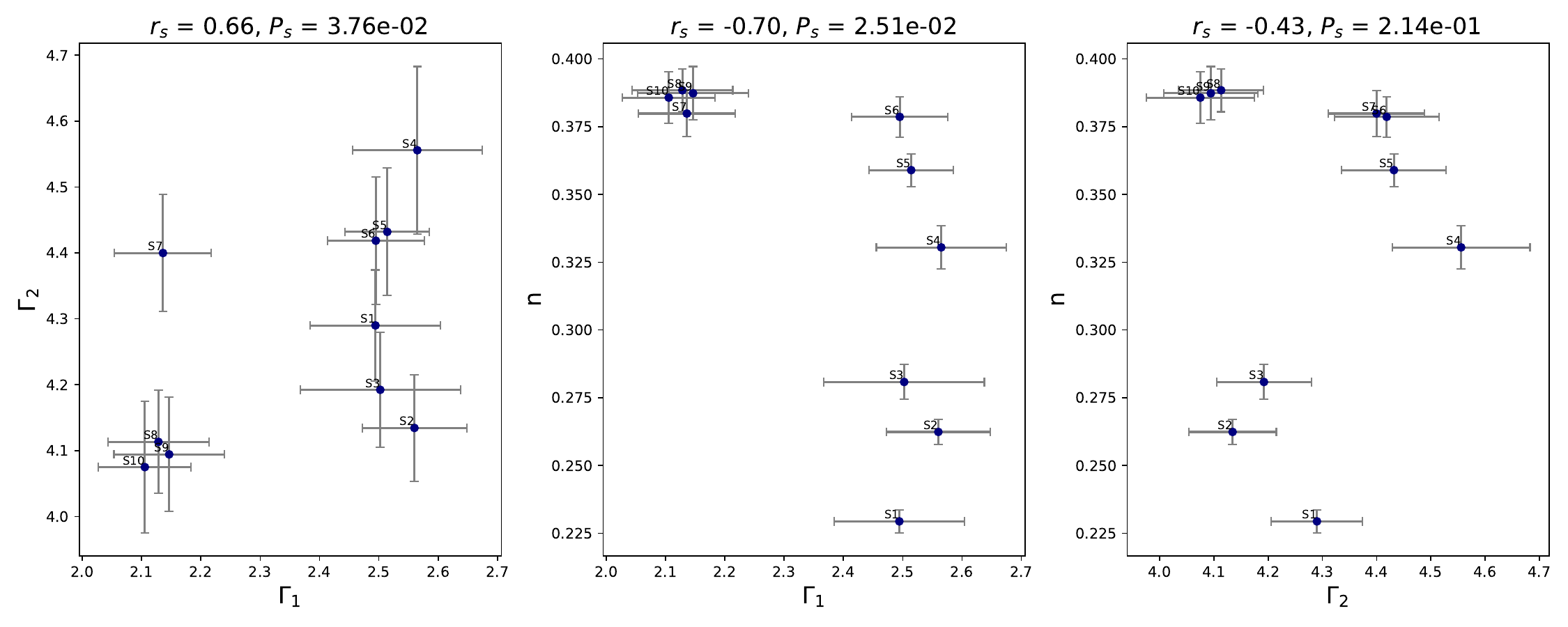}
    \caption{Correlations among the spectral parameters of the broken power-law model for Mrk\,421.}
    \label{fig:bkn_corr}
\end{figure*}

\begin{figure*}
    \centering
    \includegraphics[width=\linewidth]{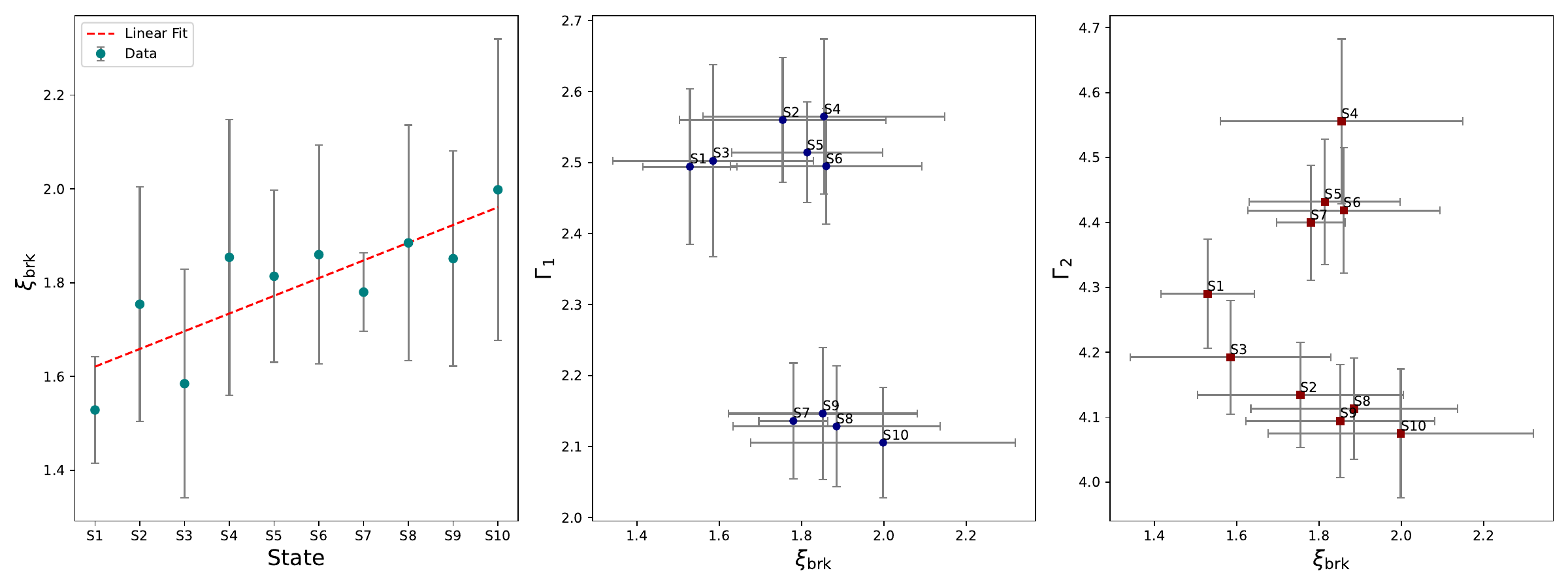}
    \caption{Variation of the break energy with flux state (S1--S10) obtained from flux resolved spectroscopy.}
    \label{fig:bkr_trends}
\end{figure*}

The variation of the broken power-law parameters across flux states (Fig.~\ref{fig:bknpow_state_trends}) 
reveals that the low-energy index ($\Gamma_1$) is distributed over two distinct levels, 
suggesting the presence of two electron populations or emission zones contributing to the observed X-ray spectrum. 
In contrast, the high-energy  index ($\Gamma_2$) shows  variation with increasing flux, while the normalization ($n$) increases monotonically from low to high states. 
This behaviour indicates that the overall X-ray spectral variability of Mrk\,421 is primarily driven by changes in the high-energy index ($\Gamma_2$), 
while the dual states seen in $\Gamma_1$ hint at intermittent transitions between two particle acceleration regimes or distinct emission regions within the jet. A gradual increase in $\xi_{\mathrm{brk}}$ with state number (see Figure \ref{fig:bkr_trends}) suggests a progressive shift of the synchrotron peak towards higher energies as the source brightens. 
The correlation plots (Fig.~\ref{fig:bkn_corr}) further confirm this interpretation.


The flux-resolved spectral analysis reveals the presence of two distinct index states, suggesting the coexistence of two particle populations within the emission region. To further investigate this behavior, we intended to perform flux and index distribution analyses. However, since the flux-resolved analysis comprises only ten data points, the sample size is insufficient to establish reliable statistical distributions. To overcome this limitation, we employed time-resolved spectroscopy, where the spectra were extracted over uniform 10\,ks intervals. This segmentation provides a larger number of spectra for long observations such as the one used in the present study, thereby allowing a more detailed examination of flux and index variability patterns.

\subsection{Time resolved spectroscopy}

The time-resolved X-ray spectroscopy for this observation (A02\_005T01\_9000000948) was previously presented in detail, together with four additional observations, in our earlier work \citep{AKBAR2025438}. In that study, the joint SXT and LAXPC20 spectra were analysed using uniform 10 ks segments to trace the spectral evolution of Mrk\,421. Since the data reduction and model-fitting procedures have already been described in Sections~\ref{reduction} and~\ref{xray_spectral}, respectively, we refer the reader to \citep{AKBAR2025438} for a complete description of the analysis. In the present work we use the spectral parameters obtained in that study for each 10 ks segment. Specifically, we adopt the broken power-law spectral parameters reported in \citep{AKBAR2025438} and use them here for comparison with the
flux-resolved results. The resulting time-resolved spectral parameters, along with their corresponding reduced-$\chi^2$
values, are listed in Table~\ref{tab:t2_bpl_logF}. It should be noted that the previous study did not compute the time-resolved flux in physical units ($\mathrm{erg\,cm^{-2}\,s^{-1}}$). Therefore, in the present work we calculate the flux for each 10 ks segment using the same spectral model (the synchrotron-convolved broken power-law model).

\subsection{Index and Flux Distributions}
\subsubsection{Observation ID:A02\_005T01\_9000000948}
\begin{figure*}
    \centering

    \begin{minipage}{0.48\textwidth}
        \centering
        \includegraphics[width=\linewidth]{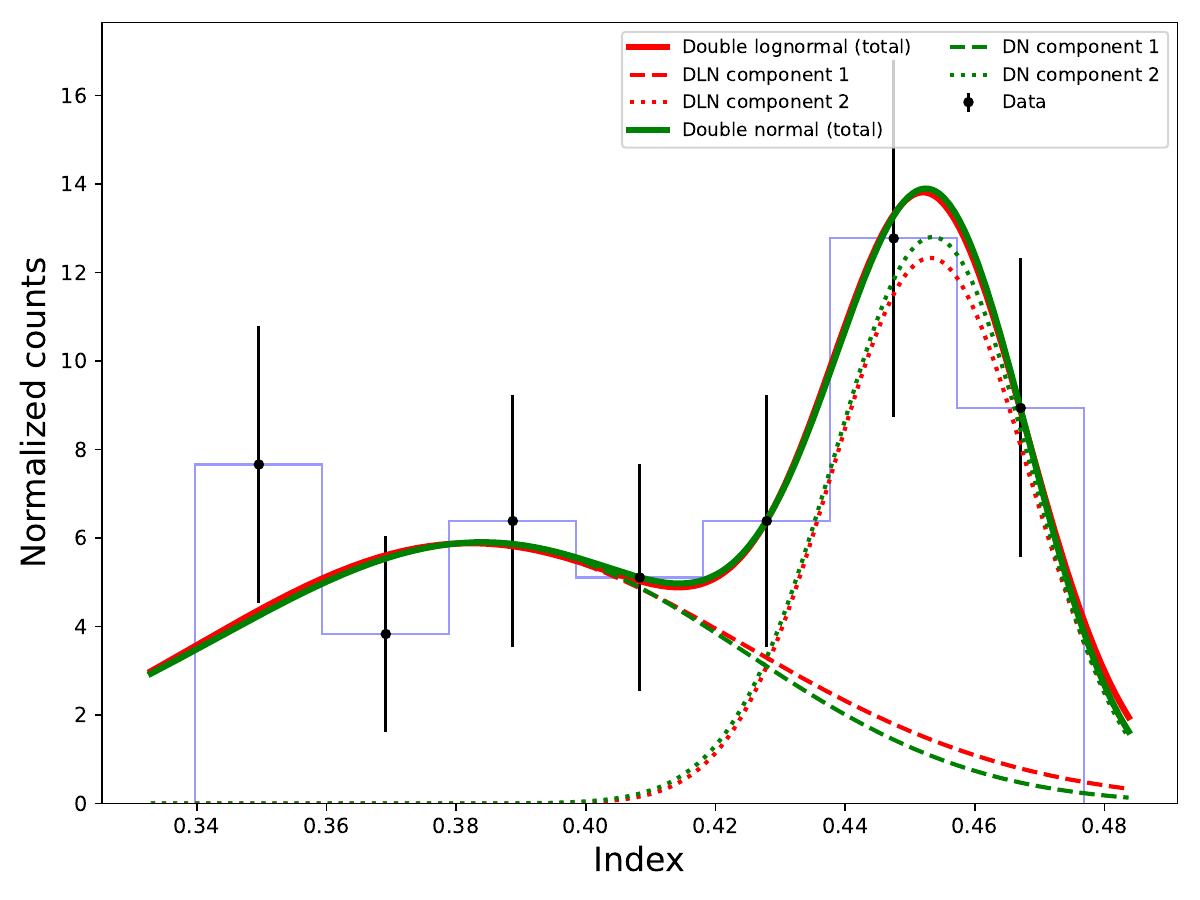}
        \label{fig:index}
    \end{minipage}
    \hfill
    \begin{minipage}{0.48\textwidth}
        \centering
        \includegraphics[width=\linewidth]{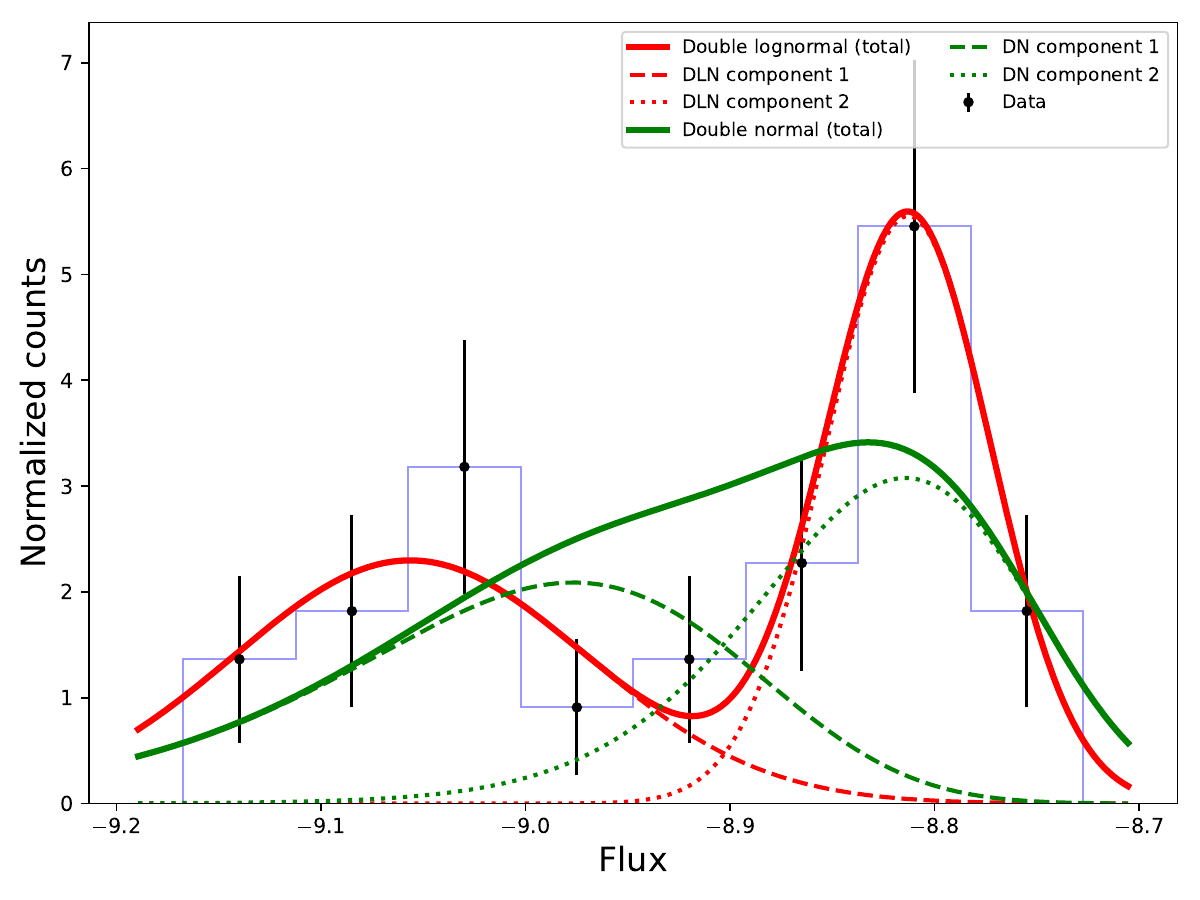}
        \label{fig:flux}
    \end{minipage}

    \caption{
Normalized histograms of the spectral index (left panel) and the X-ray flux 
($\mathrm{erg\,cm^{-2}\,s^{-1}}$) (right panel) derived from the time-resolved 
spectroscopy of the AstroSat observation A02\_005T01\_9000000948 of Mrk\,421. 
The spectra were extracted in uniform 10 ks intervals and fitted with the 
synchrotron-convolved broken power-law model.
The black markers represent the observed normalized histogram counts.
The solid red curve shows the total double lognormal (DLN) model, while the
red dashed and red dotted curves represent the two individual components of
the DLN distribution. For comparison, the solid green curve shows the total
double normal (DN) model, and the green dashed and green dotted curves denote
the two components of the DN distribution.}
    \label{fig:index_flux}
\end{figure*}

To investigate the statistical behaviour of the variability during the observation
A02\_005T01\_9000000948, we examined the distributions of the time–resolved flux and
index using the Anderson--Darling (AD) normality test together with histogram
fitting. Both the single normal and single lognormal models were unable to represent the
observed distributions, as indicated by AD statistics that exceed the 5\% critical value
(0.726). To obtain a more accurate description of the probability density function (PDF),
we constructed a normalized histogram of the logarithm of flux and index as shown in Figure~\ref{fig:index_flux} and fitted them using two-component models, namely the double lognormal and double normal functions defined in
Eqs.~\ref{eq:double_lognorm}--\ref{eq:double_norm}.

\begin{equation}
\label{eq:double_lognorm}
\begin{split}
D_{\mathrm{dLN}}(x) 
&= a\,\frac{1}{\sqrt{2\pi}\,\sigma_{1}}
   \exp\!\left[-\frac{(x-\mu_{1})^{2}}{2\sigma_{1}^{2}}\right]  \\
&\quad + (1-a)\,\frac{1}{\sqrt{2\pi}\,\sigma_{2}}
   \exp\!\left[-\frac{(x-\mu_{2})^{2}}{2\sigma_{2}^{2}}\right].
\end{split}
\end{equation}

where $a$ is the mixing fraction, $\mu_1$ and $\mu_2$ are the centroids of the two components, and $\sigma_1$ and $\sigma_2$ are their corresponding widths.

\begin{equation}
\label{eq:double_norm}
D_{\rm dN}(x) =
a\,N(x;\sigma_{1},\mu_{1}) + (1-a)\,N(x;\sigma_{2},\mu_{2}),
\end{equation}

with
\begin{equation}
N(x;\sigma,\mu) =
\frac{1}{\sqrt{2\pi}\,\sigma}
\exp\!\left[-\frac{(10^{x}-\mu)^{2}}{2\sigma^{2}}\right]
\,10^{x}\ln 10.
\end{equation}
Flux and index distribution statistics for Observation ID: A02 005T01 9000000948 are shwn in Table~\ref{tab:single_obs}.
For the flux distribution, the double log-normal model yields a reduced chi-square of $\chi^{2}_{\nu} = 1.05$, whereas the double normal representation gives $\chi^{2}_{\nu} = 6.58$ (see Table~\ref{tab:single_obs}). This indicates that the flux variability is better described by two lognormal components, a behaviour commonly associated with multiplicative variability or the presence of more than one emission state. The spectral index for the same observation also fails the normality tests. Both the double log-normal and double normal models provide fits with $\chi^{2}_{\nu} \approx 2$ (see Table~\ref{tab:single_obs}), which suggests that the index values cluster into at least two groups, possibly reflecting transitions between different spectral regimes within the source.

\subsubsection{Combined Data}
In order to assess whether the statistical behaviour inferred from Observation~ID: A02\_005T01\_9000000948 remains consistent over longer temporal baselines, we extended the analysis to two additional observations with substantially longer exposures, namely  T01\_218T01\_9000001852 and A05\_015T01\_9000002650. These observations, together with A02\_005T01\_9000000948, provide a combined temporal coverage spanning several years (2017--2019) with high-quality SXT and LAXPC20 exposures (see Table~1 of \citep{AKBAR2025438}). For each observation, the time-resolved index values were taken from our previous analysis \citep{AKBAR2025438}, while in the present study we have calculated the corresponding flux values for all segments using the same spectral model to ensure consistency. We then carried out a combined statistical analysis using all three datasets to examine whether the flux and index distribution characteristics persist across different epochs. This approach enables us to assess whether the double lognormal/normal behaviour identified in A02\_005T01\_9000000948---such as the presence of two distinct flux and spectral index states---is also manifested in other long observations, thereby providing a more comprehensive view of the long-term variability properties of Mrk\,421.

\begin{figure*}
    \centering

    \begin{minipage}{0.48\textwidth}
        \centering
        \includegraphics[width=\linewidth]{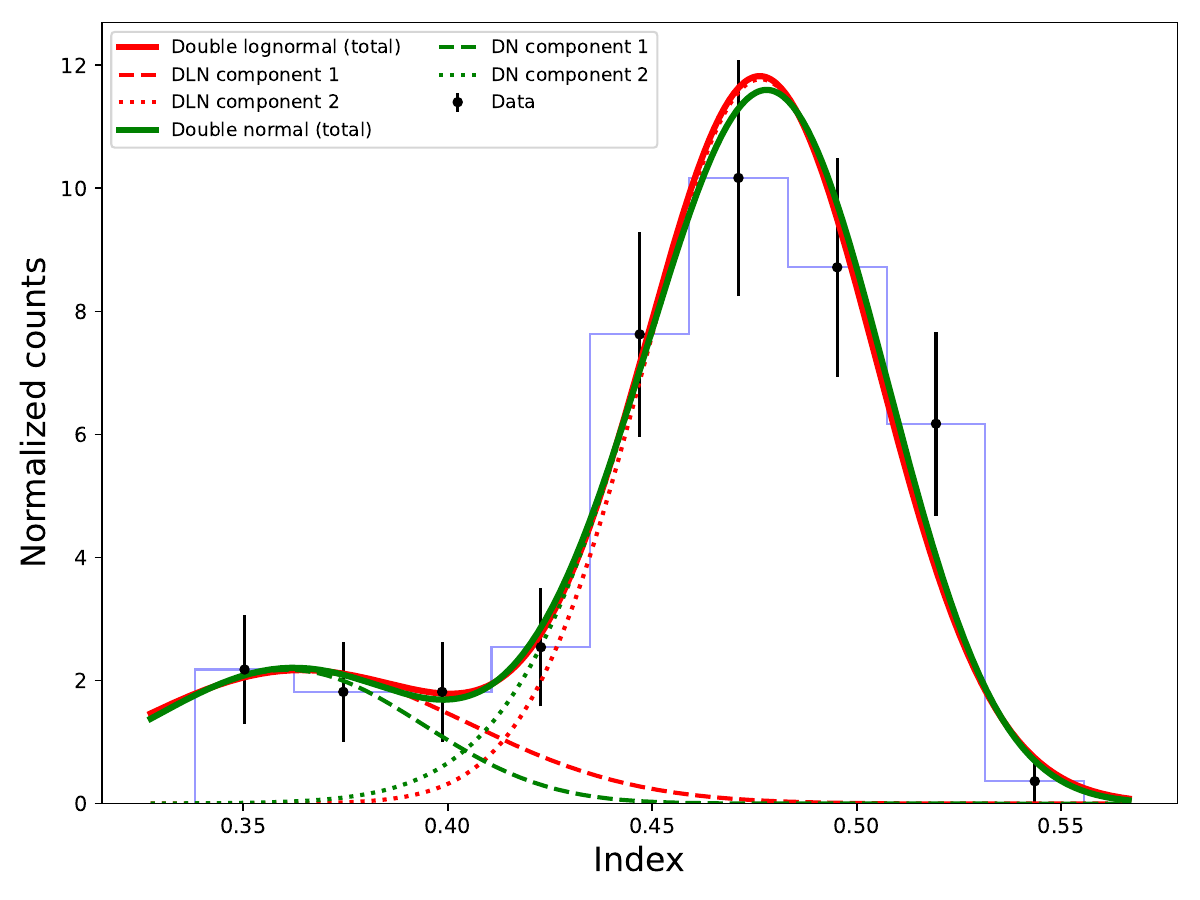}
        \label{fig:index}
    \end{minipage}
    \hfill
    \begin{minipage}{0.48\textwidth}
        \centering
        \includegraphics[width=\linewidth]{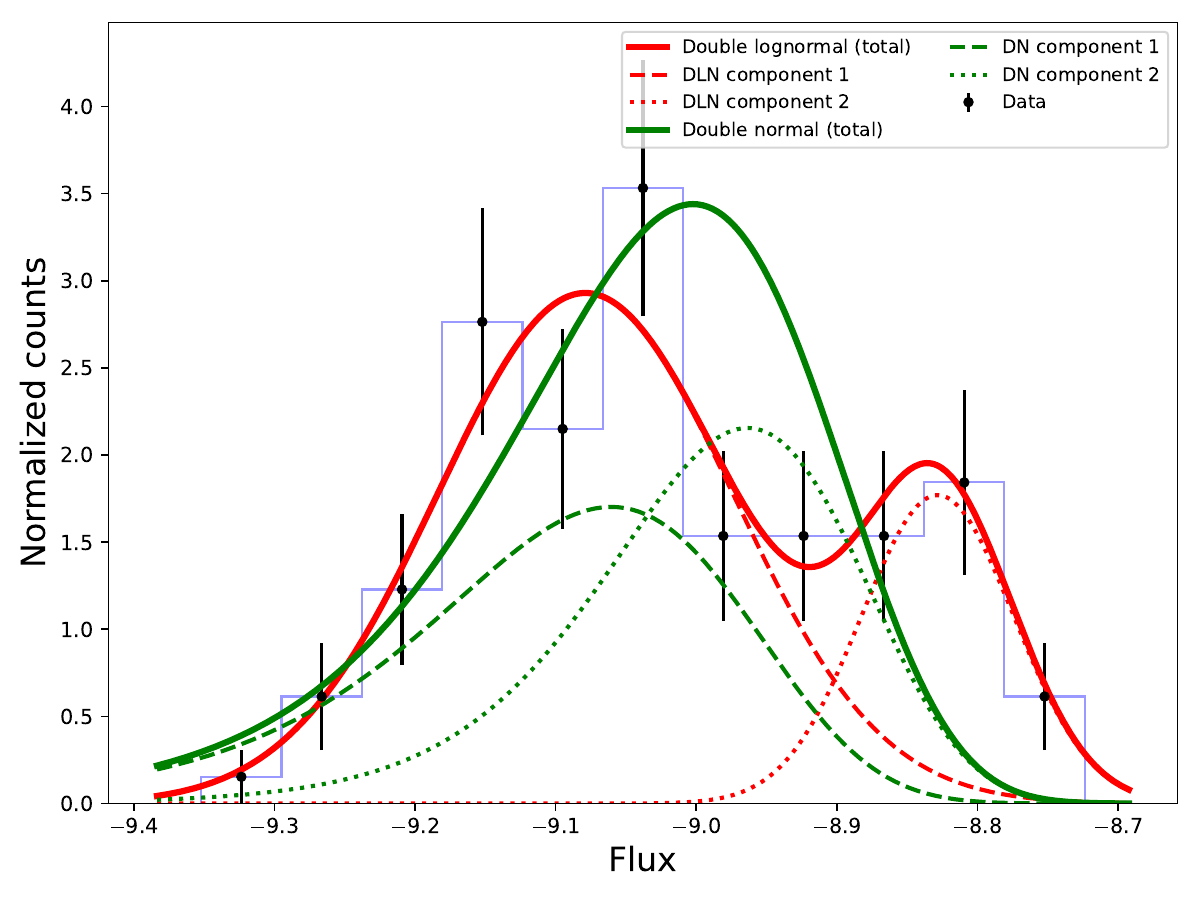}
        \label{fig:flux}
    \end{minipage}
\caption{
Combined distributions of the spectral index and X-ray flux 
($\mathrm{erg\,cm^{-2}\,s^{-1}}$) for Mrk\,421 constructed using time-resolved
spectra from multiple AstroSat observations between 2017-2019. Each
spectrum was fitted with the synchrotron-convolved broken power-law model,
and the resulting spectral parameters were used to construct the normalized
histograms.
The black markers indicate the observed histogram counts.
The solid red curve represents the total double lognormal (DLN) model,
while the red dashed and red dotted curves show its two individual
components. The solid green curve corresponds to the total double normal
(DN) model, and the green dashed and green dotted curves represent the two
components of the DN distribution.}

    \label{fig:com_index_flux}
\end{figure*}

For the combined observations, the flux and index distributions again show significant deviations from Gaussian and lognormal statistics. We constructed a normalized histogram of the logarithm of flux and index as shown in Figure~\ref{fig:com_index_flux}.  Flux and index distribution statistics for the combined observations are shown in Table~\ref{tab:combined_obs} . The combined flux distribution is well represented by a double log-normal model with $\chi^{2}_{\nu} = 0.91$, whereas the double normal model yields $\chi^{2}_{\nu} = 7.06$ (see Table~\ref{tab:combined_obs}), indicating that an additive (Gaussian-like) process is unlikely. The two lognormal components, centred at $\mu_{1}=-9.08$ and $\mu_{2}=-8.83$, point to the presence of multiple flux states, which may correspond to changes in the dominant emission region or to different levels of source activity. For the combined index distribution, both Gaussian and lognormal hypotheses are rejected by the AD tests. The double normal model provides the lower reduced chi-square ($\chi^{2}_{\nu} = 1.29$ compared to $1.50$ for the double log-normal), indicating that at least two distinct index populations are present across the observations. This behaviour is consistent with the source undergoing changes in particle acceleration or cooling conditions over time, resulting in multiple spectral states.

\begin{table*}
\centering
\caption{Flux(erg\,cm$^{-2}$\,s$^{-1}$) and index distribution statistics for Observation ID ``A02 005T01 9000000948''.}
\label{tab:single_obs}
\begin{tabular}{lcc}
\hline
\textbf{Quantity} & \textbf{Flux} & \textbf{Index} \\
\hline
Gaussian AD statistic & 1.33 & 0.74 \\
Lognormal AD statistic & 1.64 & 0.87 \\

\hline
Double log-normal: $a$ & $0.47 \pm 0.11$ & $0.57 \pm 0.33$ \\
Double log-normal: $\sigma_1$ & $0.09 \pm 0.04$ & $0.04 \pm 0.03$ \\
Double log-normal: $\mu_1$ & $-9.06 \pm 0.03$ & $0.38 \pm 0.03$ \\
Double log-normal: $\sigma_2$ & $0.04 \pm 0.01$ & $0.02 \pm 0.01$ \\
Double log-normal: $\mu_2$ & $-8.81 \pm 0.01$ & $0.45 \pm 0.01$ \\
Double log-normal: $\chi^2_\nu$ & 1.05 & 2.02 \\
\hline
Double normal: $a$ & $0.50 \pm 0.02$ & $0.56 \pm 0.33$ \\
Double normal: $\mu_1$ & $(1.00\pm0.38)\times10^{-9}$ & $2.40 \pm 0.15$ \\
Double normal: $\sigma_1$ & $(2.40\pm0.38)\times10^{-10}$ & $0.23 \pm 0.18$ \\
Double normal: $\mu_2$ & $(1.50\pm0.37)\times10^{-9}$ & $2.84 \pm 0.04$ \\
Double normal: $\sigma_2$ & $(2.40\pm0.37)\times10^{-10}$ & $0.10 \pm 0.06$ \\
Double normal: $\chi^2_\nu$ & 6.58 & 2.04 \\
\hline
\end{tabular}
\end{table*}

\begin{table*}
\centering
\caption{Flux(erg\,cm$^{-2}$\,s$^{-1}$) and index distribution statistics for the combined observations.}
\label{tab:combined_obs}
\begin{tabular}{lcc}
\hline
\textbf{Quantity} & \textbf{Combined Flux} & \textbf{Combined Index} \\
\hline
Gaussian AD statistic & 2.88 & 1.49 \\
Lognormal AD statistic & 1.11 & 2.35 \\

\hline
Double log-normal: $a$ & $0.76 \pm 0.08$ & $0.21 \pm 0.12$ \\
Double log-normal: $\sigma_1$ & $0.11 \pm 0.01$ & $0.04 \pm 0.03$ \\
Double log-normal: $\mu_1$ & $-9.08 \pm 0.02$ & $0.36 \pm 0.02$ \\
Double log-normal: $\sigma_2$ & $0.05 \pm 0.02$ & $0.03 \pm 0.00$ \\
Double log-normal: $\mu_2$ & $-8.83 \pm 0.02$ & $0.48 \pm 0.00$ \\
Double log-normal: $\chi^2_\nu$ & 0.91 & 1.50 \\
\hline
Double normal: $a$ & $0.50 \pm 0.01$ & $0.17 \pm 0.10$ \\
Double normal: $\mu_1$ & $(8.08\pm1.91)\times10^{-10}$ & $2.28 \pm 0.11$ \\
Double normal: $\sigma_1$ & $(2.30\pm0.19)\times10^{-10}$ & $0.18 \pm 0.12$ \\
Double normal: $\mu_2$ & $(1.04\pm0.19)\times10^{-9}$ & $2.99 \pm 0.03$ \\
Double normal: $\sigma_2$ & $(2.30\pm0.19)\times10^{-10}$ & $0.21 \pm 0.02$ \\
Double normal: $\chi^2_\nu$ & 7.06 & 1.29 \\
\hline
\end{tabular}
\end{table*}


\begin{table*}
\centering
\caption{Broken power-law fit parameters for the time-resolved  X-ray spectra of the S2 observation.}
\scalebox{1.0}{
\begin{tabular}{rlllll}
\hline
Time & $\Gamma_{1}$ & $\Gamma_{2}$ & norm (n) & $\log F$ & $\chi^{2}/\mathrm{dof}$ \\
 &  &  & $(\times 10^{-2})$ &(erg\,cm$^{-2}$\,s$^{-1}$)  &  \\
\hline
5000   & $2.90^{+ 0.08}_{- 0.08}$ & $ 4.70^{+ 0.16}_{- 0.15}$ & $ 11.97^{+ 0.19}_{- 0.19}$ & $ -9.131^{+ 0.008}_{- 0.008}$ & $ 1.03$ \\
15000  & $2.90^{+ 0.09}_{- 0.09}$ & $ 4.84^{+ 0.16}_{- 0.15}$ & $ 12.31^{+ 0.22}_{- 0.22}$ & $ -9.140^{+ 0.009}_{- 0.009}$ & $ 0.93$ \\
25000  & $2.98^{+ 0.07}_{- 0.07}$ & $ 4.57^{+ 0.14}_{- 0.14}$ & $ 11.48^{+ 0.17}_{- 0.17}$ & $ -9.159^{+ 0.007}_{- 0.007}$ & $ 1.04$ \\
35000  & $2.91^{+ 0.10}_{- 0.10}$ & $ 4.42^{+ 0.16}_{- 0.15}$ & $ 12.90^{+ 0.26}_{- 0.26}$ & $ -9.098^{+ 0.822}_{- 0.822}$ & $ 1.14$ \\
45000  & $2.88^{+ 0.20}_{- 0.19}$ & $ 4.32^{+ 0.16}_{- 0.15}$ & $ 12.51^{+ 0.47}_{- 0.46}$ & $ -9.104^{+ 0.019}_{- 0.019}$ & $ 0.78$ \\
55000  & $2.86^{+ 0.18}_{- 0.18}$ & $ 4.11^{+ 0.15}_{- 0.15}$ & $ 13.35^{+ 0.45}_{- 0.44}$ & $ -9.051^{+ 0.018}_{- 0.018}$ & $ 0.95$ \\
65000  & $2.74^{+ 0.09}_{- 0.08}$ & $ 4.35^{+ 0.13}_{- 0.13}$ & $ 14.64^{+ 0.23}_{- 0.23}$ & $ -9.031^{+ 0.008}_{- 0.008}$ & $ 0.95$ \\
75000  & $2.80^{+ 0.09}_{- 0.09}$ & $ 4.55^{+ 0.17}_{- 0.16}$ & $ 14.41^{+ 0.26}_{- 0.26}$ & $ -9.054^{+ 0.009}_{- 0.009}$ & $ 0.80$ \\
85000  & $2.88^{+ 0.06}_{- 0.06}$ & $ 4.31^{+ 0.13}_{- 0.12}$ & $ 13.84^{+ 0.18}_{- 0.18}$ & $ -9.058^{+ 0.007}_{- 0.007}$ & $ 1.01$ \\
95000  & $2.86^{+ 0.09}_{- 0.08}$ & $ 4.39^{+ 0.13}_{- 0.13}$ & $ 13.57^{+ 0.22}_{- 0.22}$ & $ -9.072^{+ 0.008}_{- 0.008}$ & $ 1.01$ \\
105000 & $2.68^{+ 0.17}_{- 0.16}$ & $ 4.39^{+ 0.15}_{- 0.14}$ & $ 14.28^{+ 0.38}_{- 0.37}$ & $ -9.043^{+ 0.015}_{- 0.015}$ & $ 0.85$ \\
115000 & $2.98^{+ 0.08}_{- 0.07}$ & $ 4.34^{+ 0.13}_{- 0.12}$ & $ 14.16^{+ 0.23}_{- 0.22}$ & $ -9.052^{+ 0.008}_{- 0.008}$ & $ 0.91$ \\
125000 & $2.86^{+ 0.09}_{- 0.09}$ & $ 4.48^{+ 0.14}_{- 0.14}$ & $ 14.68^{+ 0.26}_{- 0.26}$ & $ -9.043^{+ 0.009}_{- 0.009}$ & $ 0.87$ \\
135000 & $2.78^{+ 0.19}_{- 0.19}$ & $ 4.68^{+ 0.18}_{- 0.17}$ & $ 15.55^{+ 0.53}_{- 0.52}$ & $ -9.025^{+ 0.017}_{- 0.018}$ & $ 1.03$ \\
145000 & $2.74^{+ 0.16}_{- 0.15}$ & $ 4.46^{+ 0.15}_{- 0.14}$ & $ 16.52^{+ 0.44}_{- 0.44}$ & $ -8.986^{+ 0.014}_{- 0.015}$ & $ 1.04$ \\
155000 & $2.75^{+ 0.08}_{- 0.07}$ & $ 4.53^{+ 0.14}_{- 0.13}$ & $ 17.82^{+ 0.25}_{- 0.25}$ & $ -8.956^{+ 0.007}_{- 0.007}$ & $ 0.86$ \\
165000 & $2.66^{+ 0.06}_{- 0.06}$ & $ 4.37^{+ 0.12}_{- 0.11}$ & $ 19.87^{+ 0.22}_{- 0.22}$ & $ -8.897^{+ 0.006}_{- 0.006}$ & $ 1.05$ \\
175000 & $2.57^{+ 0.08}_{- 0.08}$ & $ 4.71^{+ 0.15}_{- 0.14}$ & $ 20.54^{+ 0.28}_{- 0.28}$ & $ -8.897^{+ 0.007}_{- 0.007}$ & $ 0.95$ \\
185000 & $2.75^{+ 0.06}_{- 0.05}$ & $ 4.52^{+ 0.12}_{- 0.11}$ & $ 19.83^{+ 0.22}_{- 0.22}$ & $ -8.909^{+ 0.006}_{- 0.006}$ & $ 0.99$ \\
195000 & $2.60^{+ 0.06}_{- 0.06}$ & $ 4.48^{+ 0.11}_{- 0.11}$ & $ 21.98^{+ 0.23}_{- 0.23}$ & $ -8.853^{+ 0.006}_{- 0.006}$ & $ 1.05$ \\
205000 & $2.71^{+ 0.08}_{- 0.08}$ & $ 4.50^{+ 0.13}_{- 0.12}$ & $ 21.85^{+ 0.33}_{- 0.33}$ & $ -8.863^{+ 0.008}_{- 0.008}$ & $ 1.05$ \\
215000 & $2.66^{+ 0.08}_{- 0.08}$ & $ 4.32^{+ 0.11}_{- 0.11}$ & $ 21.61^{+ 0.30}_{- 0.30}$ & $ -8.854^{+ 0.008}_{- 0.008}$ & $ 1.02$ \\
225000 & $2.65^{+ 0.19}_{- 0.18}$ & $ 4.06^{+ 0.12}_{- 0.12}$ & $ 22.11^{+ 0.66}_{- 0.64}$ & $ -8.819^{+ 0.017}_{- 0.018}$ & $ 1.01$ \\
235000 & $2.57^{+ 0.08}_{- 0.08}$ & $ 3.88^{+ 0.11}_{- 0.10}$ & $ 22.44^{+ 0.30}_{- 0.30}$ & $ -8.794^{+ 0.008}_{- 0.008}$ & $ 1.08$ \\
245000 & $2.49^{+ 0.08}_{- 0.08}$ & $ 3.87^{+ 0.11}_{- 0.10}$ & $ 23.37^{+ 0.31}_{- 0.31}$ & $ -8.769^{+ 0.008}_{- 0.008}$ & $ 1.00$ \\
255000 & $2.58^{+ 0.06}_{- 0.06}$ & $ 3.97^{+ 0.10}_{- 0.10}$ & $ 22.22^{+ 0.23}_{- 0.23}$ & $ -8.808^{+ 0.006}_{- 0.006}$ & $ 1.00$ \\
265000 & $2.49^{+ 0.06}_{- 0.06}$ & $ 4.23^{+ 0.11}_{- 0.10}$ & $ 22.23^{+ 0.22}_{- 0.22}$ & $ -8.823^{+ 0.006}_{- 0.006}$ & $ 0.94$ \\
275000 & $2.43^{+ 0.07}_{- 0.07}$ & $ 4.30^{+ 0.11}_{- 0.11}$ & $ 22.32^{+ 0.25}_{- 0.25}$ & $ -8.825^{+ 0.006}_{- 0.006}$ & $ 0.92$ \\
285000 & $2.45^{+ 0.09}_{- 0.08}$ & $ 4.39^{+ 0.12}_{- 0.12}$ & $ 22.54^{+ 0.30}_{- 0.30}$ & $ -8.827^{+ 0.008}_{- 0.008}$ & $ 0.97$ \\
295000 & $2.39^{+ 0.08}_{- 0.08}$ & $ 4.19^{+ 0.11}_{- 0.10}$ & $ 25.21^{+ 0.32}_{- 0.32}$ & $ -8.760^{+ 0.008}_{- 0.008}$ & $ 0.97$ \\
305000 & $2.39^{+ 0.16}_{- 0.15}$ & $ 4.30^{+ 0.13}_{- 0.12}$ & $ 23.60^{+ 0.51}_{- 0.51}$ & $ -8.804^{+ 0.014}_{- 0.014}$ & $ 0.99$ \\
315000 & $2.26^{+ 0.17}_{- 0.17}$ & $ 4.44^{+ 0.13}_{- 0.13}$ & $ 24.48^{+ 0.54}_{- 0.54}$ & $ -8.783^{+ 0.014}_{- 0.015}$ & $ 0.82$ \\
325000 & $2.23^{+ 0.08}_{- 0.08}$ & $ 4.41^{+ 0.12}_{- 0.12}$ & $ 23.02^{+ 0.28}_{- 0.28}$ & $ -8.806^{+ 0.007}_{- 0.007}$ & $ 1.05$ \\
335000 & $2.20^{+ 0.06}_{- 0.06}$ & $ 3.96^{+ 0.10}_{- 0.10}$ & $ 24.43^{+ 0.25}_{- 0.25}$ & $ -8.736^{+ 0.006}_{- 0.006}$ & $ 0.86$ \\
345000 & $2.21^{+ 0.06}_{- 0.06}$ & $ 4.27^{+ 0.11}_{- 0.10}$ & $ 25.30^{+ 0.24}_{- 0.24}$ & $ -8.748^{+ 0.005}_{- 0.005}$ & $ 0.95$ \\
355000 & $2.78^{+ 0.28}_{- 0.26}$ & $ 4.47^{+ 0.15}_{- 0.15}$ & $ 21.74^{+ 1.10}_{- 1.05}$ & $ -8.875^{+ 0.026}_{- 0.027}$ & $ 0.72$ \\
365000 & $2.42^{+ 0.07}_{- 0.07}$ & $ 4.62^{+ 0.12}_{- 0.11}$ & $ 23.45^{+ 0.27}_{- 0.27}$ & $ -8.825^{+ 0.007}_{- 0.007}$ & $ 0.93$ \\
375000 & $2.26^{+ 0.07}_{- 0.07}$ & $ 4.89^{+ 0.13}_{- 0.12}$ & $ 24.41^{+ 0.25}_{- 0.25}$ & $ -8.811^{+ 0.006}_{- 0.006}$ & $ 0.83$ \\
385000 & $2.24^{+ 0.13}_{- 0.12}$ & $ 4.81^{+ 0.15}_{- 0.14}$ & $ 23.83^{+ 0.39}_{- 0.40}$ & $ -8.818^{+ 0.010}_{- 0.010}$ & $ 0.86$ \\
395000 & $2.32^{+ 0.19}_{- 0.19}$ & $ 4.84^{+ 0.16}_{- 0.15}$ & $ 22.38^{+ 0.58}_{- 0.57}$ & $ -8.851^{+ 0.016}_{- 0.016}$ & $ 0.90$ \\
\hline
\end{tabular}
}
\label{tab:t2_bpl_logF}
\end{table*}

\section{Summary and Discussion}\label{summary}

Mrk\,421 is an HBL blazar and one of the brightest blazars observed in the X-ray band. The source was observed during 3--8 January 2017 with the SXT and LAXPC20 instruments onboard \textit{AstroSat}. In the present analysis, the SXT data were used in the 0.8--7.0\,keV range and the LAXPC20 data in the 3.0--30.0\,keV range, providing broad X-ray spectral coverage over the synchrotron component. The observation offers one of the longest uninterrupted LAXPC exposures of this source, enabling a detailed investigation of its short- and long-term variability characteristics. The 100\,s binned LAXPC20 light curve shows pronounced flux variations throughout the observation window, capturing rapid fluctuations. The LAXPC20 data, complemented by simultaneous SXT coverage, allowed us to investigate the evolution of the spectral shape with changing flux.
To quantify the variability, we computed the fractional variability amplitude, $F_{\rm var}$, for both the SXT and LAXPC20 light curves. We obtained $F_{\rm var}=0.2103 \pm 0.0051$ for SXT and $F_{\rm var}=0.3159 \pm 0.0056$ for LAXPC20, confirming strong and significant variability during the observation. The LAXPC20 light curve exhibits a general rise in flux, reaching a peak count rate of 122.94\,counts\,s$^{-1}$. Such pronounced variability on timescales of a few hundred seconds points to compact emission regions and rapid particle acceleration episodes within the jet.

To investigate the evolution of the X-ray spectrum with brightness, we performed flux-resolved spectroscopy by dividing the background-subtracted 100\,s binned LAXPC20 light curve into distinct flux intervals, which were grouped into ten flux states (S1--S10).
The resulting LAXPC20+SXT spectra were fitted with the synchrotron convolved BPL and LP model.  We noted that the BPL model provided comparatively better fit than LP model in all the flux states, highlighting the presence of spectral break.  The BPL fits show that the low-energy particle index, 
$\Gamma$1, clusters around two distinct values across the flux states, indicating intermittent transitions between two electron populations. In contrast, the high-energy  index, $\Gamma$2, shows random changes with flux, demonstrating that the spectral variability is primarily driven by changes in the high-energy electron distribution. 
Above the break, we observe synchrotron photons from the highest-energy electrons. The distribution of these electrons are sensitive to acceleration, cooling and escape time scales. Small fluctuation in these electrons can produce large irregular changes in the slope of high energy electrons. 
Further, the break energy ($\xi_{brk}$) increases gradually from low to high flux states, implying a shift of the synchrotron peak toward higher energies as the source brightens. This is a very well-known behaviour in Mrk\,421 and other HBLs,  where flare states are often associated with a harder spectrum and higher peak energy \citep{2004ApJ...601..759T, 2016MNRAS.458...56W}.
Physically, this trend implies that during brighter states, particle acceleration becomes more efficient, enabling electrons to reach higher characteristic energies. As a result, the synchrotron cooling break shifts to higher energies, a trend we clearly detect through the rising $\xi_{brk}$ values. The correlated trends among $\Gamma$1,  $\Gamma$2, normalization, and $\xi_{brk}$ further support the picture of a dynamically evolving electron population, in which spectral changes are closely tied to flux variability.

 The low energy index of BPL (below $E_{brk}$) does not vary smoothly with flux; instead it clusters around two discrete values across all flux states. Physically, low-energy X-rays  in Mrk\,421 are produced by lower energy electrons (below $E_{brk}$), which directly trace the injection spectrum. Since these electrons are less sensitive to rapid radiative cooling, their spectral properties reflect conditions in relatively stable regions of the jet, such as zones with slower changes in magnetic configuration or particle acceleration efficiency. The presence of  two preferred $\Gamma_1$ values suggests that the system transitions between two distinct spectra states. 
 
 To further probe whether this variability arises from a single emission state or multiple ones, we performed time-resolved spectroscopy of Mrk\,421 for Observation ID A02\_005T01\_9000000948 by segmenting the data into uniform 10\,ks intervals. The resulting LAXPC20+SXT spectra were fitted using the  broken power-law model, allowing us to study spectral evolution on intermediate timescales. The fits reveal substantial variability in all  spectral parameters ($\Gamma_1, \Gamma_2$, $\xi_{brk}$). 
 By computing the flux for each segment, we are able to directly relate these spectral variations to the temporal behaviour of the source.

 To examine the statistical nature of this variability, we analyzed the distributions of time-resolved flux and particle index. We noted that both Gaussian and log-normal models fail to describe the observed distributions, as confirmed by the AD test. Instead, histogram fitting showed that the flux distribution is best represented by a double log-normal function, suggesting that the variability is multiplicative in origin but arises from two distinct flux components. Similarly, the index distribution exhibited a two-component structure, with a double Gaussian (double normal) model providing a significantly better fit than any single-component alternative. These findings align with the results of our flux-resolved analysis and support the presence of two dominant spectral states. To assess whether this double lognormal/normal behaviour is persistent over longer timescales, we extended our statistical analysis to include two additional long-exposure AstroSat observations from 2017–2019. The combined dataset shows a clear departures from both Gaussian and single lognormal statistics. As with the single observation, the flux distribution is strongly favoured by a double log-normal model, with a substantially lower reduced chi-square than the double normal fit. The centroids of the two lognormal components indicate the presence of two dominant flux states, suggesting that source routinely alternates between distinct levels of activity. The combined index distribution remains inconsistent with single-component models, and is best approximated by a double normal distribution, again implying the existence of two spectral index states across different epochs. 
 
These findings are consistent with previous studies of Mrk\,421 and other blazars. For instance, \citet{2020MNRAS.491.1934K} analyzed 16 years of RXTE-PCA data for Mrk\,421 and Mrk\,501 and found that both sources exhibit double lognormal flux distributions and double normal particle index histograms, strongly supporting the existence of two spectral states. The theoretical foundation for the connection between index variations and flux distributions is provided by \citet{2018MNRAS.480L.116S}, who showed that Gaussian perturbations in the particle acceleration timescale can naturally give rise to lognormal flux variability. In their stochastic acceleration model, Gaussian scatter in the acceleration (or escape) timescale leads to a skewed, lognormal distribution in the flux, while the particle index distribution remains approximately Gaussian.
Indeed, as \citet{2020MNRAS.491.1934K} argue, “a Gaussian distribution of index produces a lognormal distribution in flux; therefore, a double Gaussian distribution of index in Mrk\,501 and Mrk\,421 indicates that their flux distributions are probably double lognormal.” Similar two-state signatures have also been reported in other studies of blazars across different energy bands and timescales \citep{2025PhRvD.111l3052S, 2025MNRAS.539.2185M}.
The presence of a double lognormal flux distribution and a double normal index distribution in our data has important implications. It suggests that Mrk\,421's X-ray emitting region can switch between two distinct physical states, potentially governed by changes in particle acceleration efficiency, magnetic field configuration, or turbulence within the jet. When the acceleration timescale transitions between two Gaussian regimes, the source switches between “low” and “high” flux states, each associated with its own typical particle index. In statistical terms, this leads to the double lognormal flux distribution and double normal index distribution observed in the histograms. A potential concern in interpreting two-state spectral behaviour in blazars is whether such signatures arise from intrinsic physical state changes or from methodological choices such as temporal binning, flux selection, or model degeneracy. In this work, we find that the same two-state behaviour in Mrk\,421 is recovered independently through flux-resolved spectroscopy and through time-resolved statistical distributions of flux and spectral index. The agreement between these approaches, which probe variability in fundamentally different ways, indicates that the observed state-switching is not an artifact of a particular analysis method or timescale selection. Instead, it reflects a persistent property of the source across both short and long timescales.

\section{Acknowledgements}
 SA is thankful to the MOMA for the MANF fellowship (No.F.82-27/2019(SA-III)). ZS is supported by the Department of Science and Technology, Govt. of India, under the INSPIRE Faculty grant (DST/INSPIRE/04/2020/002319). SA and ZS  express  gratitude to the Inter-University Centre for Astronomy and Astrophysics (IUCAA) in Pune, India, for the support and facilities provided.

\bibliographystyle{elsarticle-harv} 
\bibliography{sample631}





 
 \section{Appendix}

\end{document}